\documentclass[prl,twocolumn,showpacs,preprintnumbers,amsmath,amssymb]{revtex4}
\usepackage{dcolumn}
\usepackage{multirow}
\usepackage{epsfig}
\usepackage[english]{babel}

\begin{document}

\bibliographystyle{try}
\topmargin 0.1cm

\newcounter{univ_counter}
\setcounter{univ_counter} {0}
\addtocounter{univ_counter} {1}
\edef\HISKP{$^{\arabic{univ_counter}}$ } \addtocounter{univ_counter}{1}

\title{Is chiral symmetry broken and or restored in high-mass light baryons\,?}

\author{
E.~Klempt~\\
Helmholtz--Institut f\"ur Strahlen-- und Kernphysik, Universit\"at
Bonn, Germany}

\begin{abstract}
Based on a thorough comparison of the nucleon and $\Delta$
excitation spectrum with models we show that parity doublets
observed in the mass spectra do not entail the consequence that
highly excited $N$ or $\Delta$ resonances are insensitive to chiral
symmetry breaking. Instead, the mechanism of mass generation in
excited states is suggested to be the same as for the baryon ground
states: the mass is assigned to fluctuating gluon fields and their
strong attraction. In excited baryons, the field energy has to be
integrated over a larger volume, and the total mass increases. Thus,
also the additional mass of resonances, the excitation energy, is
generated by spontaneous breaking of chiral symmetry.
\end{abstract}
\maketitle

SU(3) symmetry and the conjecture that mesons and baryons are
composed of constituent quarks \cite{GellMann:1964nj,Zweig:1964jf}
paved the path to an understanding of the particle zoo.  A
constituent light-quark mass of about 350\,MeV was required to
reproduce the masses of ground-state baryons; the
$N$--$\Delta(1232)$ mass splitting and the pattern of negative- and
positive-parity excited baryons was interpreted as an effect of a
QCD hyperfine interaction between these constituent quarks
\cite{Isgur:1977ef,Isgur:1978xj}. However, low-energy approximations
of QCD \cite{Weinberg:1978kz} lead to the Gell-Mann-Oakes-Renner
relation \cite{GellMann:1968rz} which assigns a mass of a few MeV to
light (current) quarks. The mass gap between current and constituent
quarks is interpreted by spontaneous breaking of the chiral symmetry
expected for nearly massless quarks
\cite{Nambu:1960tm,Goldstone:1961eq}. An important consequence is
the large mass gap between chiral partners: the masses of the
nucleon, with spin-parity ${\mathtt J^{\mathtt P}}=1/2^+$, and its
chiral partner $N_{1/2^-}(1535)$, with spin-parity $\mathtt
J^{\mathtt P}=1/2^-$ and mass $M=1535$\,MeV, differ by about
600\,MeV.

In the higher mass region, parity doublets are observed and often,
nucleon and $\Delta$ resonances of given spin and parity form a
quartet of mass-degenerate states.  Most convincing examples are the
light-quark baryons in the third and forth resonance region (see
Table \ref{tab:second}). In both regions, two mass-degenerate
spin-isospin quartets with ${\mathtt J^{\mathtt P}}=1/2^{\pm}$ and
${\mathtt J^{\mathtt P}}=3/2^{\pm}$, respectively, can be identified
(in the first and second column). In the third region, a ${\mathtt
J}=5/2^{\pm}$ parity doublet of nucleon resonances, in the forth
region, a ${\mathtt J^{\mathtt P}}=5/2^{\pm}$ parity doublet of
$\Delta$ resonances can be recognized. A parity partner of
$N_{5/2^+}(2000)$ is missing. $N_{1/2^+}(1440)$ and
$N_{1/2^-}(1535)$ in the 2$^{\rm nd}$ resonance region are not
really mass-degenerate, $\Delta_{3/2^+}(1232)$ and $N_{3/2^-}(1520)$
have no close-by parity partner; the 1$^{\rm st}$ and 2$^{\rm nd}$
resonance regions do not yet belong to the highly excited states.

The observation of parity doublets has led to the conjecture that
chiral symmetry might be effectively restored in highly excited
baryons \cite{Glozman:1999tk}. The mass generation mechanism in
excited hadrons is, according to Glozman \cite{Glozman:2008vg}, very
different compared to the mechanism in the lower-mass states. In the
latter states, the mass is supposed to be driven by chiral symmetry
breaking in the vacuum, by the quark condensate. For highly excited
states, the quark condensate is believed to be almost irrelevant and
the mass of resonances within a parity doublet could have a chirally
symmetric origin.

The conjecture of chiral symmetry restoration of highly excited
hadrons has been worked out in a number of papers, we quote a few
recent reviews \cite{Jaffe:2004ph}. Particularly exciting would be
the possibility to track the transition from constituent quarks to
current quarks by precise measurements of the masses of excited
hadron resonances \cite{Bicudo:2009cr}. A weak attraction between
parity partners in the 2\,GeV mass region -- as suggested by
phenomenology -- can possibly be interpreted as onset of a regime in
which chiral symmetry is restored \cite{Klempt:2002tt}. This
interpretation depends, of course, crucially on the assumption that
chiral symmetry breaking plays no role in the high-mass part of the
hadron excitation spectrum.

\begin{table}[pb]
\caption{\label{tab:second}Light-quark nucleon ($N$) and $\Delta$
resonances in the 1$^{\rm st}$, 2$^{\rm nd}$, 3$^{\rm rd}$, and
4$^{\rm th}$ resonance region (rr). The spin-parities ${\mathtt
J}^{\mathtt P}$ of resonances are given as subscripts.
}\begin{center}
\begin{tabular}{lcccc}
\hline\hline\\[-2.2ex]
\ ${\mathtt J^{\mathtt P}}=\ 1/2^{\pm}$\qquad & $3/2^{\pm}$ & $5/2^{\pm}$ &$7/2^{\pm}$ &rr\\ \hline\\[-2.2ex]
& $\Delta_{3/2^+}(1232)$ && &1$^{\rm st}$ \\\hline\\[-2.2ex]
$N_{1/2^-}(1535)$ & $N_{3/2^-}(1520)$ && &\multirow{2}{5mm}{2$^{\rm nd}$}\\
$N_{1/2^+}(1440)$ &  &  & \\[0.3ex]\hline\\[-2.2ex]
$N_{1/2^-}(1650)$ & $N_{3/2^-}(1700)$ & $N_{5/2^-}(1675)$ &&\\
$N_{1/2^+}(1710)$ & $N_{3/2^+}(1720)$ & $N_{5/2^+}(1680)$ &&\multirow{2}{5mm}{3$^{\rm rd}$}\\
$\Delta_{1/2^-}(1620)$ & $\Delta_{3/2^-}(1700)$ &&&\\
$\Delta_{1/2^+}(1750)$ & $\Delta_{3/2^+}(1600)$ &&&\\\hline\\[-2.2ex]
$N_{1/2^-}(1885)^1$ & $N_{3/2^-}(1875)^1$ & ?&? &   \\
$N_{1/2^+}(1880)^1$ & $N_{3/2^+}(1900)$ & $N_{5/2^+}(2000)$ & $N_{7/2^+}(1990)$&\multirow{2}{5mm}{4$^{\rm th}$}\\
$\Delta_{1/2^-}(1900)$ & $\Delta_{3/2^-}(1940)$ & $\Delta_{5/2^-}(1930)$ &? & \\
$\Delta_{1/2^+}(1910)$ & $\Delta_{3/2^+}(1920)$ &
$\Delta_{5/2^+}(1905)$ &$\Delta_{7/2^+}(1950)$&\\[0.3ex]\hline\hline\\[-1.5ex]
\end{tabular}
{\it $^1$States not reported in \cite{Amsler:2008zzb} but
observed in the\\[-0.5ex] Bonn-Gatchina multichannel partial wave analysis
\cite{Sarantsev:2010}.}
\end{center}
\end{table}
If chiral symmetry is at work in highly excited baryons, a few
additional states must exist which are indicated by question marks
in Table \ref{tab:second}. Candidates for these additional states
are $N_{5/2^-}(2200)$, $N_{7/2^-}(2190)$, and
$\Delta_{7/2^-}(2200)$, respectively. The three states are separated
from their parity partners by about 220\,MeV or $\delta M^2=
0.92$\,GeV$^2$ which is suspiciously close to the string tension
characterizing the slope of baryon Regge trajectories
(1.06\,GeV$^2$). Hence the question must be answered if the absence
(or ``wrong" mass) of some states -- expected in scenarios with
chiral symmetry restoration -- is due to lacking experimental
information, or if we can understand the reason why some resonances
have parity partners and others not.

In this letter we discuss a dynamical origin of the occurrence of
parity doublets in the mass spectrum of mesons and baryons and show
that parity doublets in high-mass hadrons do not need to signal
chiral symmetry. On the contrary, parity doublets could be the
consequence of chiral symmetry breaking in an extended volume. The
conclusions are derived from a thorough comparison of the
experimental mass spectrum of nucleon and $\Delta(1232)$ resonances
\footnote{According to \cite{Arndt:2003}, the number of observed
states could be much smaller than the number given in
\cite{Amsler:2008zzb}. We use the full spectrum of a recent review
\cite{Klempt:2009pi}, for reasons given in \cite{Sarantsev:2010}.}
with the conjecture that chiral symmetry is restored
\cite{Glozman:2008vg}, with quark model predictions
\cite{Capstick:1986bm,Loring:2001kx}, with the Skyrme model
\cite{Mattis:1984dh}, and with predictions of an analytically
solvable ``gravitational" theory simulating QCD
\cite{deTeramond:2005su,Karch:2006pv,Forkel:2007cm,Brodsky:2010ev}
which is defined in a five-dimensional Anti-de Sitter (AdS) space
embedded in six dimensions. Here, a special variant
\cite{Forkel:2008un} of AdS/QCD is used.

Quark models are the traditional approach to hadron spectroscopy.
The pattern of low-mass states is, perhaps, reasonably well
described but the models fail in important details. First, the
number of predicted states below, e.g., 2.2\,GeV is excessively
large, much larger than the number of experimentally known states.
This is called the problem of {\it missing resonances}. Second, the
predicted mass pattern is not really adequate. For the $\Delta$
excitations, quark models predict a shell structure roughly
compatible with the pattern of a harmonic oscillator: a
positive-parity ground state, two negative-parity states, followed
by a group of positive-parity states, then negative parity, positive
parity, $\cdots$. Data do not exhibit the even-odd staggering of
masses predicted by quark models. Instead, $\Delta$ resonances with
even and odd parities cluster at approximately equidistant
mass-square values. A $\chi^2$ test of quark models
\cite{Capstick:1986bm,Loring:2001kx} versus experiment thus gives
modest agreement only, in spite of a significant number of free
parameters. A Skyrme model \cite{Mattis:1984dh} uses fewer
parameters and the agreement with data is worse (see Table
\ref{tab:comp}. A detailed comparison can be found in
\cite{Klempt:2010du}).

\begin{table}[pt]
\caption{\label{tab:comp}Comparison of models with data. The number
of parameters is given and  a ``quality" factor. For $Q=2.5$\% the
rms model deviation from experiment is 50\,MeV at 2\,GeV
corresponding to $\sim20$\% of the natural widths. In the high-mass
region, a large number of states is predicted by quark models. The
smallest mass difference is chosen for the comparison.}
\begin{center}
\begin{tabular}{cccc} \hline\hline\\[-2.5ex]  Ref. & \ \ $N_p$ & $Q$
\\\hline\\[-2.5ex]
\cite{Capstick:1986bm} & \quad 7 \quad &\quad  $ (\delta M/M) = 5.6$\% \\[0.2ex]
\cite{Loring:2001kx} & \quad   5 \quad &\quad  $ (\delta M/M) = 5.1$\%\\[0.2ex]
\cite{Mattis:1984dh} & \quad  2 \quad &\quad  $ (\delta M/M) = 9.1$\%\\[0.2ex]
\cite{Forkel:2008un} & \quad  2 \quad &\quad  $(\delta M/M) = 2.5$\%\\[0.2ex]
\hline\hline
\end{tabular}\\[1.5ex]
\end{center}
\renewcommand{\arraystretch}{1.0}
\end{table}

In meson spectroscopy, a large number of resonances comes as well in
nearly mass degenerate parity doublets, however with important
exemptions: Mesons like $f_2(1270)$ and $a_2(1320)$ with ${\mathtt
J^{\mathtt{PC}}=2^{\mathtt{++}}}$, $\omega_3(1670)$ and
$\rho_3(1690)$ with ${\mathtt J^{\mathtt{PC}}=3^{\mathtt{--}}}$,
$f_4(2050)$ and $a_4(2040)$ with ${\mathtt
J^{\mathtt{PC}}=4^{\mathtt{++}}}$, none of these states falling onto
the leading Regge trajectory has a mass-degenerate spin-parity
partner. These are mesons in which the orbital angular momentum
${\mathtt L}$ and the total quark spin ${\mathtt S}$ are aligned to
give the maximal ${\mathtt J}$ and which have the lowest mass in
that partial wave. Their chiral partners have considerably higher
masses: $\eta_2(1645)$ and $\pi_2(1670)$ (${\mathtt
J^{\mathtt{PC}}=2^{\mathtt{-+}}}$), $h_3(2045)$ and $b_3(2035)$
(${\mathtt J^{\mathtt{PC}}=3^{\mathtt{+-}}}$), $\eta_4(2320)$ and
$\pi_4(2250)$ (${\mathtt J^{\mathtt{PC}}=4^{\mathtt{-+}}}$),
respectively \cite{Bugg:2004xu}. A graphical illustration is given
in Fig.~1 of \cite{Afonin:2006wt} and Fig.~57 of
\cite{Klempt:2007cp}.

The meson spectrum is compatible with a simple formula derived in
AdS/QCD \cite{Forkel:2007cm}
\begin{eqnarray}
M^2 =& a\cdot (\mathtt{L+N+}1/2)& {\rm for\ mesons}
\label{N_ADS}
\end{eqnarray}
with $a=1.14\left[{\mathrm GeV^2}\right]$ as Regge slope parameter.
The total angular momentum $\mathtt J$ (the spin of the resonance)
does not appear in Eq.~(\ref{N_ADS}): the orientation of the total
quark spin ${\mathtt{S}}$ along the orbital angular momentum
${\mathtt{L}}$ and the spin-spin interaction -- leading to spin
singlet- and spin-triplet mesons -- have no significant impact on
the meson mass, at least not for mesons with $\mathtt J\neq 0$.
Scalar and pseudoscalar mesons are governed by additional forces (by
four-quark and meson-meson interactions and, respectively, by their
coupling to gluons leading to the $\rm U_A(1)$ anomaly); their
masses are not well reproduced by eq.~(\ref{N_ADS}). But otherwise,
the formula is very successful; in particular it reproduces the
correct pattern where parity partners should be observed and where
not.

In \cite{Glozman:2010rp} it is argued that formation of the
spin-parity partners of mesons on the leading Regge trajectory could
be suppressed by angular momentum barrier factors. A reanalysis of
the reaction $\bar pp\to \pi\eta\eta$ in flight
\cite{Anisovich:2000ut} was performed and a weak indication claimed
\cite{Glozman:2010rp} for the possible existence of the missing
$4^{-+}$ state $\eta_4(1950)$ at about 1.95\,GeV. However, the
weakness of the signal is certainly not enforcing any
interpretation. Nevertheless, the suppression in $\bar pp$ formation
of spin-parity partners of mesons on the leading Regge trajectory is
certainly an argument which reduces the weight of their
non-observation. Hence we go back to baryons.

The conjecture that chiral symmetry is restored gives an
interpretation of the mass degeneracy (within a resonance region)
along vertical lines in Table \ref{tab:second}; the mass degeneracy
along the horizontal lines is still accidental. Here, AdS/QCD is
much more powerful. In the variant \cite{Forkel:2008un} it predicts
\begin{eqnarray}
M^2 =& a\cdot (\mathtt{L+N}+3/2)& -b\,\alpha_D \label{M_ADS}
\label{M_ADS}
\end{eqnarray}
a formula which has been suggested before on an empirical basis
\cite{Klempt:2002vp}. The baryon Regge trajectory requires a
slightly softer slope, $a=1.06\left[{\mathrm GeV^2}\right]$.
$\mathtt L$ is the total intrinsic orbital angular momentum and
$\mathtt{N}$ the radial quantum number. Quark models of baryons have
two oscillators (like any bound three-body problem), and hence four
quantum numbers $l_1,l_2,n_1,n_2$. Eq.~(\ref{M_ADS}) contains
$\mathtt L$ and $\mathtt{N}$ only. AdS/QCD predicts therefore a much
smaller number of states. $\alpha_D$ is the fraction of {\it good
diquarks} in the baryon, of diquarks with zero spin and isospin. The
good-diquark fraction can be calculated from standard quark-model
wave functions. It is 1/2 in the nucleon, 1/4 in the
$N_{1/2^-}(1535)/N_{3/2^-}(1520)$ spin doublet, and it is assumed to
be 1/2 (1/4) for all spin-1/2 nucleon resonances in SU(6) 56-plets
(70-plets). In spin-3/2 nucleon and in $\Delta$ excitations, there
are no good diquarks. The coefficient $b=1.46\left[{\mathrm
GeV^2}\right]$ gives the best fit. Without this term, the agreement
between AdS/QCD and data is considerably worse for negative-parity
spin-1/2 nucleons and for all spin-3/2 nucleons even if different
slopes for negative parity baryons are admitted.

The appearance of the orbital angular momentum $\mathtt L$ in
Eqs.~(\ref{N_ADS}) and (\ref{M_ADS}) is intriguing. A discussion has
developed if the use of non-relativistic concepts is legitimate for
the dynamics of quarks in highly excited states
\cite{Glozman:2009bt}. Even a constituent quark mass of 350\,MeV is
small compared to the mass of a highly excited hadron, and it can be
argued that relativistic effects must be huge. On the other hand,
Teramond and Brodsky \cite{deTeramond:2008ht} have shown that bound
states in QCD with arbitrary spin and intrinsic angular momentum can
be mapped onto string modes in AdS/QCD with defined angular
momentum, and that the classification of states with AdS/QCD quantum
numbers is hence legitimate. For the moment we put aside doubts
concerning the applicability of a non-relativistic notion in hadron
spectroscopy and show that it is at least a useful concept.

In Table~\ref{tab:second}, the fifth line exhibits a spin triplet of
states. It is natural to assign an intrinsic orbital angular
momentum $\mathtt L=1$ and total quark spin $\mathtt S=3/2$ which
couple to a total angular momentum $\mathtt{J}=1/2, 3/2, 5/2$. The
three masses are similar, the spin-orbit interaction is obviously
small, and also mixing with other states having the same quantum
numbers does not have a significant impact on the masses. (The three
states in the sixth line do not form a spin triplet but a spin
singlet with $\mathtt{L}=0$ and a spin doublet with $\mathtt{L}=2$.)
Likewise, there are two spin quartets of nucleon and $\Delta$
excitations with $\mathtt{L}=2$, listed in line 10 and 12 of
Table~\ref{tab:second}. The spin doublet $\Delta_{1/2^-}(1620)$ and
$\Delta_{3/2^-}(1700)$ (line 7) has $\mathtt L=1$ and intrinsic spin
${\mathtt S}=1/2$. For $\mathtt L=1, \mathtt S=3/2$, symmetry
arguments enforce $\tt{N}=1$, and these states are found in the
second but last line of Table~\ref{tab:second}. Their isotopic
companions, $N_{1/2^-}(1885)$ and $N_{3/2^-}(1875)$, can be
interpreted as radial excitations of $N_{1/2^-}(1535)$ and
$N_{3/2^-}(1520)$, respectively. The resonances
$\Delta_{1/2^+}(1600)/N_{1/2^+}(1710)$ -- and $\Delta_{3/2^+}(1750)$
if it exists \cite{Sarantsev:2010} -- can be interpreted as first
($\tt{N}=1$)/second ($\mathtt{N}=2$) radial excitations of the
respective ground states.

AdS/QCD reproduces the masses of all 44 $N$ and $\Delta$ resonances
remarkably well using just two parameters, considerably better than
other models (see Table \ref{tab:comp}). One parameter in
Eq.~(\ref{M_ADS}) is related to confinement, the second one accounts
for hyperfine effects. It reduces the size of the nucleon by a
fraction which depends on $\alpha_D$. The precision of the mass
calculation is by far better than quark model predictions even
though the latter have a significantly larger number of parameters.
Obviously, AdS/QCD catches the correct variable which governs the
excitation spectrum. This is surprising since $\mathtt L$ and
$\mathtt S$ are not defined in a relativistic situation but only
$\mathtt J$.

The decisive variable is size. In AdS/QCD, the size of a hadron is
limited, either by a so-called {\it hard wall} beyond which the wave
function has to vanish or by a repelling dilaton field which
increases quadratically with the extension of the wave function and
which is called {\it soft wall}. In most AdS/QCD approaches to
hadron spectroscopy, a soft wall limits the extent of the wave
function. In \cite{Forkel:2008un}, the second term in
Eq.~(\ref{M_ADS}) is constructed to reduce the size of the system as
a function of the good diquark content. A nucleon is thus smaller
than the $\Delta(1232)$.

Why is size important for the mass of a resonance? We first discuss
the nucleon mass. In massless QCD, their is no scale in the QCD
Lagrangian and hence one should expect the nucleon mass to vanish.
But this is obviously wrong. Chiral symmetry is spontaneously
broken, and the nucleon mass receives not only contributions from
quarks; the fluctuating color-electric and color-magnetic fields
$\bf E$ and $\bf B$ carry most of the nucleon mass
\cite{Wilczek:2006eg}:
\begin{eqnarray}
M_N = < N\mid -\frac{9 \alpha_s}{4 \pi} ({\bf B}^2-{\bf
E}^2)+\hspace{-2mm}\sum_{\rm flavors} m_i\bar\psi_i\psi_i \mid N>\;.
\label{divergence}
\end{eqnarray}
The fields need to be integrated over the over the hadron volume.
AdS/QCD predicts that the size of a hadron plays the decisive role
for its mass. With increasing size, chiral symmetry is spontaneously
broken in an extended volume. If the field strengths in
Eq.~(\ref{divergence}) are approximately uniformly spread over the
volume, the stored field energy increases with volume and hence the
(squared) mass increses. Loosely speaking, the constituent quark
mass increases to about 1/3 of the mass of the resonance. The
dynamical degrees of freedom in hadron spectroscopy are thus not
constituent quarks having a defined rest mass, a large kinetic
energy, and some residual interaction; instead, constituent quark
masses are seen to be ill-defined. In excited hadrons they adopt
much larger values, typically 1/3 of the hadron mass. This
justifies, a posteriori, the use of orbital angular momentum
$\mathtt L$ and quark spin $\mathtt S$ in hadron spectroscopy,
quantities which the data demand and which are key quantities in
AdS/QCD even though they are not defined in relativistic situations.

Finally we ask if angular momentum barriers may be responsible for
the non-observation of the chiral partners of, e.g.,
$N_{7/2^+}(1990)$ and $\Delta_{7/2^+}(1950)$. In $\pi N$ elastic
scattering, this is indeed the case. In scattering, an angular
momentum ${\mathtt L}=3$ is required to form a $7/2^+$ resonance;
for $7/2^-$, ${\mathtt L}=4$ is needed. Hence there is still a way
to escape the conclusions offered here and to rescue the conjecture
of chiral symmetry restoration in highly excited hadrons. This
hideout can be closed in photoproduction experiments: a $7/2^+$
resonance requires a $E_4 ^+$ or $M_4 ^+$ amplitude, a $7/2^-$
resonance a $E_3^-$ or $M_3^-$ amplitude. There is no kinematical
factor which would suppress production of $7/2^-$ resonances
compared to $7/2^+$ resonances. Photoproduction experiments can thus
be of decisive importance to clarify the dynamics of highly excited
hadrons.

Summarizing, we have shown that size is the quantity which governs
the mass of light-quark baryons. Chiral symmetry breaking,
responsible for the proton mass, seems to generate the mass of
excited states as well. Likely, there is no chiral symmetry
restoration in highly excited baryons. We propose that constituent
quarks, if introduced, should not be considered to have a defined
rest mass. Instead, their rest mass increases with increasing baryon
mass, and the mass of the three constituent quarks accounts for the
essential part of the resonance mass. This conjecture provides a
natural explanation of the long-standing miracle that the naive
non-relativistic quark model is surprisingly successful. We have
argued that the search for missing states expected in
parity-doublets scenarios should be performed in production and not
in formation experiments. Photoproduction of nucleon and $\Delta$
resonances with $\mathtt{J}^\mathtt{P}$=$5/2^{\pm}$ and $7/2^{\pm}$,
expected at masses in the 1.9 to 2.3\,GeV mass range, can be
decisive to clarify if chiral symmetry is broken or restored in
high-mass baryon resonances.

\begin{acknowledgments}
I would like to thank S. J.~Brodsky, H. Forkel, U.~G. Mei\ss ner, V.
Metag, G.~F. de~Teramond, and W. Weise  for elucidating discussions
and S. J.~Brodsky, H. Forkel, and U.~G. Mei\ss ner for a critical
reading of the manuscript. Financial support from the Deutsche
Forschungsgemeinschaft (DFG) within SFB/TR16 is kindly acknowledged.
\end{acknowledgments}

\end{document}